\documentclass{article}

\usepackage{PRIMEarxiv}

\usepackage[utf8]{inputenc} 
\usepackage[T1]{fontenc}    
\usepackage{hyperref}       
\usepackage{url}            
\usepackage{booktabs}       
\usepackage{amsfonts}       
\usepackage{nicefrac}       
\usepackage{microtype}      
\usepackage{lipsum}
\usepackage{fancyhdr}       
\usepackage{graphicx}       
\graphicspath{{media/}}     

\title{Personal Data Visualisation on Mobile Devices: A Systematic Literature Review
\thanks{\textit{\underline{Citation}}: 
\textbf{Authors. Title. Pages.... DOI:000000/11111.}} 
}

\author{
  Yasmeen anjeer alsehhhi, Mohamed Abdelrazek \\
   \\
  Deakin Univeristy \\
  \texttt{\{Yasmeen, Mohamed\}yanjeeralshehhi@deakin.edu.au@mohamed.abdlerazek@deakin.edu.au} \\
   \And
  Alessio Bonti \\
  Deakin University \\
  \texttt{a.bonti@deakin.edu.au} \\
}

\begin{document}
\maketitle

\begin{abstract}
Personal data cover multiple aspects of our daily life and activities, including health, finance, social, Internet, Etc. Personal data visualisations aim to improve the user experience when exploring these large amounts of personal data and potentially provide insights to assist individuals in their decision making and achieving goals. People with different backgrounds, gender and ages usually need to access their data on their mobile devices. Although there are many personal tracking apps, the user experience when using these apps and visualisations is not evaluated yet. There are publications on personal data visualisation in the literature. Still, no systematic literature review investigated the gaps in this area to assist in developing new personal data visualisation techniques focusing on user experience. In this systematic literature review, we considered studies published between 2010 and 2020 in three online databases. We screened 195 studies and identified 29 papers that met our inclusion criteria. Our key findings are various types of personal data, and users have been addressed well in the found papers, including health, sport, diet, Driving habits, lifelogging, productivity, Etc. The user types range from naive users to expert and developers users based on the experiment's target. However, mobile device capabilities and limitations regarding data visualisation tasks have not been well addressed. There are no studies on the best practices of personal data visualisation on mobile devices, assessment frameworks for data visualisation, or design frameworks for personal data visualisations. 

\end{abstract}

\keywords{Personal data visualization \and mobile visualization\and  tracking visualization}

\section{Introduction}
Personal analytics, personal visual analytics \cite{r28}, Personal data visualisation, quantified self's \cite{r12}, or self-tracking are interchangeable terms\cite{r6}\cite{r7}. It is used to define the science of analytical reasoning displayed using visual representations to explore personal context internally such as  goals, skills and preferences and externally that are related to physical constraints such as devices or social influence\cite{r28}. Lee et al. divided personal information into five groups: body information, psychological state and traits, activity, environmental property status, social interaction and other life logging aspects such as tracking financial data \cite{r6}. Individuals may be interested in tracking various conditions \cite{r5}. However, the main self-tracking motivation is health conditions using different applications through mobile devices \cite{r5}, and wearable devices such as smartwatches \cite{r5}. Adopting mobile devices to track personal data became one of the open research areas \cite{r2}. Personal data visualisation has been investigated by Choe et al. in which quantified selfers have recorded their experiences in tracking their data, this paper has only summarised the number of visualisation charts used and how frequently each of these types has been used \cite{r5}. Oh et al. have developed a study that investigated the UX issues for tracking personal data and found that simple charts can not be understood by everyone due to graph literacy, on the other hand, it is required to develop advanced data visualisation to enrich expert users experience when exploring their data \cite{r6}. Katz et al have specifically investigated the effectiveness of diabetes commercial mobile applications in terms of data visualisation and other aspects\cite{r7}. However, this was limited research in which it covered only one specific type of personal data to be visualised and tracked. The provided reviews were positive in understanding the presented information but not sufficient context information and interaction with data visualisation. 

Still, there are gaps in personal data visualisation to be adopted on mobile devices. There are open research areas related to the smart phones' capability and compatibility to adopt data visualisation charts, tasks and interactions. Another research area is the guidelines for developing complete, correct and consistent data visualisation on mobile devices. Additionally, there is no detailed systematic literature review of the state of the art and state of practice regarding personal data visualisations.

In this paper, we introduce a systematic literature review (SLR) on personal data visualisation. The SLR addresses the following research questions:
\begin{itemize}
    \item What personal data does end-users want to track?
    this research question aims to collect and show the different kinds of tracked personal data so we know which of these we should do more research on.
    \item What are the different backgrounds and skills of personal data visualisation end users?
   As personal visualisations are targeting non-expert audience, we need to understand to know who are they and what works for them and what dimensions did existing studies consider.
    \item What are the advantages and disadvantages of adopting mobile devices in tracking and visualising personal data? 
    We need to investigate and understand mobile devices pros and cons to adopt personal data visualisation focusing on both hardware and software.
    \item What are the visualisation techniques and visualisation tasks covered in the literature in personal data visualisation?
    The target of this question is to acknowledge the adopted graphs, tasks and interaction by the end users and the provided applications.
    
\end{itemize}

The rest of the paper is structured as follows: first, we discuss the related work done in personal and mobile data visualisation research areas. Second, we present the methodology we followed in conducting the SLR. Third, we present a summary of the results we found. Finally, we provide a detailed discussion on the findings from the SLR process.
\section{Related Work}
There are existing research and experiments related to personal data visualisation and adoption by users. However, the number of systematic literature reviews developed to cover this research field is limited. We found one SLR paper covered the design criteria to develop data visualisation published in 2015 \cite{r39}, and another two review papers developed in the area of evaluating software visualisation and learning analytic visualisation \cite{r36},\cite{r37}. The recent SLR that we found published in 2020 covered the type of health data for elderly people \cite{r38}.

Out of the reviews mentioned above, there is only one review that is related to our topic \cite{r38}. It reviewed 95 papers that were published between 2009 and 2019. The purpose of the review was to assess the proposed technology in terms of its easiness and allowance of entering health data and evaluating the appropriateness of the information presented with data visualisation. There are various data identified to track health data, including emotion, sleep, pain and time. Smartphones are the top devices used by older people to visualise their health data. Although this review claimed that information visualisation is straightforward in terms of understanding and interpretation of a displayed image, there are limited papers that describe the characteristics of good data visualisation. There were two main methods identified to evaluate mobile health technology, either quantitative or mixed, of both quantitative and qualitative \cite{r38}. 
In terms of data visualisation evaluation, Merino et al. conducted a literature review in which they reviewed 181 papers out of 387 papers published in SOFTVIS/VISSOFT conferences \cite{r36}. They divided the evaluation methods into two groups traditional and non-traditional. The researchers found three ways: conducting interviews with participants or developers to gain hypotheses to be tested in the experiment, developing questionnaires to reaa ch a large number of participa,nts and recordisessionsions to evaluate user performance. There are three methods classified as non-traditional to evaluate software visualisation eye tracking to capture how participants see the elements in visualisation, log analysis to investigate how participants navigate visualisations, and emotion cards to help participants measurably report their feelings.  Although these methods are listed to be effective to evaluate software visualisation, yet three main barriers are limiting the evaluation of these tools: 
\begin{itemize}
    \item The lack of ready to use an evaluation tool 
\item The lack of benchmark to compare with the developed tool
\item The difficulty to find the right type of participants to share their knowledge 
\end{itemize}

In terms of visualisation techniques, two reviews agreed on classifying visualisation techniques into traditional and modern techniques\cite{r39}, \cite{r37} While traditional visualisation could be developed using a spreadsheet such as a bar chart, line chart, box chart, pie chart. The higher-level visualisation requires special tools to create charts like geo charts (Google maps), time charts, running race charts, time logs, spiral displays, cluster maps and component planes. In terms of modern visualisation techniques, art and technology are combined to show energy consumption. For example, a floor plan can be presented, and red spots are presented in different areas of the floor to show power consumption\cite{r39}.
Designing a visualisation interface is another challenge reviewed in 2015 by covering 22 studies to develop a list of visualisation design guidelines to show energy consumption\cite{r39}. In their review, the researcher found that guidelines should be grouped into functional and non-functional criteria. The functional criteria involve the type of information displayed in the visualisation, techniques, and modes: the non-functional criteria are hardware and software integration, portability, and extensibility. Based on the findings, there are five recommendations developed based on the design requirement. An example of these recommendations is if the visualisation should be portable and displayed on different platforms, the recommendation was to focus on the scalability, flexibility and accessibility\cite{r39}. 

However, there are some missing challenges related to the type of users, visualisation techniques that can be used to show personal data, and mobile devices' ability to adopt this kind of visualisation. 
\section{Methodology}
\label{Methodlogy}
In order to conduct this systematic literature review, the researcher has followed kitchenham’s guidelines \cite{r33} to evaluate the relevant data by developing a protocol that includes research questions,  the inclusion and exclusion criteria, quality assessment, data extraction and identifying the selected studies. 
\subsection{Research Questions}
In this SLR, we have focused on the following research questions:
\begin{itemize}
\item SLR RQ1: What personal data do users prefer to view on their mobile phone?  Health? Personal development? Utility usage? Financial spending? Lifestyle monitoring?
\item SLR RQ2: What are the groups of users primarily interested in personal data visualisation?
\item SLR RQ3: To what extent are mobile devices capable of adopting personal visualisation? In terms of efficiency, scalability, memory, user experience, Etc.
\item SLR RQ4: What are the existing techniques in the literature to visualise personal data on mobile devices?
\item SLR RQ5: What is the feedback on the available applications/techniques that support personal visualisation? 
\item SLR RQ6: what are the challenges to design a visualisation interface in mobile devices? 
\end{itemize}

\subsection{Inclusion and exclusion criteria}
The researcher has adopted the following string to be applied as a search query to be used in the following search engines:
\textbf{("mobile data visualisation") OR ("personal analytics" AND "data visualisation" AND "mobile"). }
The search engines are Google Scholar, IEEE Explore, and ACM digital library, including peer-reviewed articles.
Inclusion criteria:
\begin{itemize}
\item It should be in the English language 
\item It should relate to the topic scope
\item It should be within the years from 2010 to 2020 
\item The full text should be available
\item It should be a peer-reviewed journal or conference article
\end{itemize}

\subsection{Selection Process}
\label{Selection process}
{In order to have a well-built selection process, we screened the results of the search string, using the inclusion and exclusion criteria provided previously. After that, we read the titles and abstracts of the publication to check the eligibility. Finally, we filled Table 1 to filter out the paper used in the systematic literature review.}

\begin{table}
\caption{Selection Process}
\label{tab:1}       
\begin{tabular}{lll}
\hline\noalign{\smallskip}
Title&Publication Venue &Reasons of inclusion or exclusion  \\
\noalign{\smallskip}\hline\noalign{\smallskip}
----- & ----- & ----- \\
----- & ----- & ----- \\
\noalign{\smallskip}\hline
\end{tabular}
\end{table}

\subsection{Study Quality Assessment}
\label{SQA}
In addition to the inclusion and exclusion criteria, the papers selected will undergo a quality assessment process to ensure their valuable contribution to the SLR. Based on the search we have done using the queries search mentioned above, the results can be classified into quantitative and qualitative studies. The following criteria are used to evaluate the quality of the papers generally \cite{r31},
\begin{itemize}

\item Does the paper have a description of the research method? 
\item Does the paper describe an explicit research question/goal/purpose? 
\item Does the paper describe the motivation for the research question(s)? 
\item Does the paper discuss limitations or validity? 
\item Does the paper describe the context of the research? 
\item Does the paper describe data collection? 
\item Does the paper describe data analysis? 
\item Does the paper describe sampling or selection of the study object(s)?
\item Does the paper present any data?
\end{itemize}

\subsection{Data Extraction}
\label{DE}
Usually data extraction needs a second reviewer to establish inter-rater reliability. Thus, the co-authors performed data extraction on random samples of the primary study to cross-check the results. Table 2 shows the data types to be collected: 
\begin{table}
\caption{Data Extraction}
\label{tab:2}       
\begin{tabular}{lllll}
\hline\noalign{\smallskip}
 Reference&Study Parameters&Focus&Methodology&Findings \\
\noalign{\smallskip}\hline\noalign{\smallskip}
-----& ----- & -------& ----- & ---- \\
\noalign{\smallskip}\hline
\end{tabular}
\end{table}

\subsection{Selected Studies}
Based on the inclusion criteria, out of 194 papers, only 25 papers have met the criteria. Additionally, we added three more papers we identified while reviewing the 25 papers. Figure 1 shows the process of selecting the used papers in this SLR.
\begin{figure*}[h]
  \centering
  \includegraphics[width=0.75\textwidth]{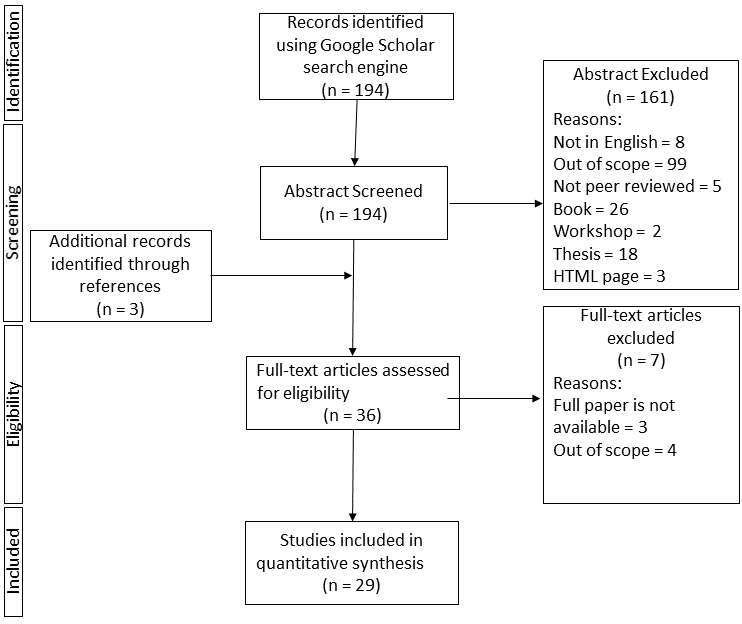}
  \caption{The Process of Selecting Studies \& [Source], via Research gate website, From Moher D, Liberati A, Tetzlaff J, Altman DG, The PRISMA Group (2009). Preferred Reporting Items for Systematic Reviews and MetaAnalyses.}
  \label{fig:1} 
\end{figure*}

The twenty-eight selected research papers have been published at 25 different venues; IEEE Transaction on visualisation and Computer Graphics has published 4 out of these 28 papers. Another two papers have been published through IEEE Transactions on Computer Graphics and Applications. In addition to IEEE Transactions, the Proceedings of the SIGCHI Conference on Human Factors in Computing Systems have published two research studies. While the mentioned venues have published more than one paper, the other venues in Table 3 have published 1 of the research study related to our topic.
\begin{table}
\caption{Venues of selected papers}
\label{tab:3}       
\begin{tabular}{ll}
\hline\noalign{\smallskip}
 Venue Name&Counts \\
\noalign{\smallskip}\hline\noalign{\smallskip}
2015 Eighth International Conference on Mobile Computing\& and Ubiquitous Networking (ICMU)  &     1 \\
    Behaviour  Information Technology & 1 \\
    Computers in Human Behaviour & 1 \\
    Economy and Society & 1  \\
    eHealth 360°,Springer & 1 \\
    Human–Computer Interaction & 1 \\
    IEEE 16th International Conference on Smart City & 1 \\
    IEEE Computer Graphics and Applications & 2 \\
    IEEE International Conference on \&Computer and Information Technology \\Ubiquitous Computing and Communications & 1\\
IEEE transactions on visualisation and computer graphics & 4 \\
International Journal of Mobile Human \& and Computer Interaction & 1\\
International Journal on Advances in Intelligent System & 1\\
International society for optics and photonics &1\\
Journal of biomedical informatics &1\\
Journal of Comparative Research\& in Anthropology and Sociology & 1\\
Journal of the Association for \& Information Science and Technology & 1 \\
Journal of Visual Languages and Computing & 1\\
Proceedings of SIGRAD 2015 & 1 \\
Proceedings of the 2018 CHI Conference \& on Human Factors in Computing Systems & 1 \\
Proceedings of the 22nd International\& Academic Mindtrek Conference &1 \\
Proceedings of the ACM on Interactive, Mobile,\& Wearable and Ubiquitous Technologies & 1 \\
Proceedings of the SIGCHI Conference\& on Human Factors in Computing Systems & 2 \\
Societies & 1\\
Big data & 1\\
Total & 29\\\\
\noalign{\smallskip}\hline
\end{tabular}
\end{table}

\section{Results}
We now detail and group the results of the research questions that have been found in the literature. The following sections will present the results based on five aspects: types of self-tracked data, user profile, target mobile devices, visualisation tasks and techniques, and user feedback. Firstly, we will show the types of personal data tracked and the types of users targeted or included in the found studies. This information is presented in Tables 4 and 5. Secondly, we will present the various devices included in the studies as mobile devices and elaborate their benefits and challenges to adopting self-tracking and data visualisation; tables 6 and 7 present these two aspects. Thirdly, we will explain the visualisation tasks and techniques in the selected research papers and their adoption in different self-tracking applications. This information is presented in Tables 8 and 9. Finally, we will discuss the feedback given by users in terms of UX design, application functionality and essential features. 
\vfill
\begin{table}
\caption{Tracked Personal Data }
\label{tab:4}       
\begin{tabular}{ll}
\hline\noalign{\smallskip}
 Tracked Personal Data&Total \\
\noalign{\smallskip}\hline\noalign{\smallskip}
Activity - Fitness& 4  \\
Diet & 2\\
Driving habits &1 \\
Energy consumption & 1 \\
Environmental property statue &2 \\
Financial & 2 \\
Geo location &1 \\
Health - Body information & 7\\
Life logging &1 \\
Productivity & 2 \\
Psychological data and traits & 2\\
Sleeping pattern &1 \\
Smartphone usage & 1\\
Social connectedness & 2 \\
Not mentioned & 17 \\
Total & 46 \\ 
\noalign{\smallskip}\hline
\end{tabular}
\end{table}

\begin{table}
\caption{Type of Users}
\label{tab:6}       
\begin{tabular}{ll}
\hline\noalign{\smallskip}
 Type of Users&Total\\
\noalign{\smallskip}\hline\noalign{\smallskip}
-Developer & 1\\
Diabetes users& 2 \\
Expert in data visualisation &1\\
Novice users & 2\\
Not specified  & 3\\
Public users  & 6 \\
Students  &1\\
Not mentioned & 13\\
Total &  29\\ \\
\noalign{\smallskip}\hline
\end{tabular}
\end{table}

\begin{table}
\caption{Devices}
\label{tab:7}       
\begin{tabular}{ll}
\hline\noalign{\smallskip}
 Devices&Total\\
\noalign{\smallskip}\hline\noalign{\smallskip}
Large display & 1\\
Mobile devices & 4 \\
Smart phones & 3\\
Tablets & 1 \\
Wearable devices & 4 \\
Not specified & 14\\
Total & 31\\
\noalign{\smallskip}\hline
\end{tabular}
\end{table}

\begin{table}
\caption{Mobile Benefits \ Challenges }
\label{tab:7}       
\begin{tabular}{ll}
\hline\noalign{\smallskip}
 Pros and Cons&Total\\
\noalign{\smallskip}\hline\noalign{\smallskip}
Benefits & 6\\
Challenges & 4\\
\noalign{\smallskip}\hline
\end{tabular}
\end{table}

\begin{table}
\caption{visualisation Types }
\label{tab:8}       
\begin{tabular}{ll}
\hline\noalign{\smallskip}
 visualisation Types&Total\\
\noalign{\smallskip}\hline\noalign{\smallskip}
Animation visualisation & 1\\
Info graphic & 2\\
Line Graph & 2\\
Mapping visualisation & 1 \\
Multivariate & 1\\
Pie chart & 1\\
Time Series Data & 1\\
Bar chart & 1 \\
Table & 1 \\
Not mentioned & 20 \\
\noalign{\smallskip}\hline
\end{tabular}
\end{table}

\begin{table}
\caption{visualisation tasks }
\label{tab:10}       
\begin{tabular}{ll}
\hline\noalign{\smallskip}
 visualisation Tasks&Total\\
\noalign{\smallskip}\hline\noalign{\smallskip}
Analytics tasks\\ (select – encode – reconfigure – elaborate -filter -connect) & 2\\
Auto -generation visualisation type-recommended& 1\\
Change colour scheme and design - recommended  & 1\\
Custom- Create their own novel -recommended  & 2\\
Read-only& 2\\
Not mentioned & 21\\
\noalign{\smallskip}\hline
\end{tabular}
\end{table}
\subsection{Self-Tracked Personal Data}
\label{STPD}

Table 4 shows the number of papers that indicated various types of personal data. Please note that the total number exceeds 28 because some studies considered multiple self-tracked data. Thus, there are ten papers out of 28 that have involved various types of tracked personal data. In addition, while 6 out of 10 papers have directly stated the types of tracked personal data, the other four papers have mentioned personal data as the type of data to be tracked using developed applications or prototypes.

As Swan stated \cite{r14}that personal data are related to an individual's daily life aspects, 6 of the found papers have listed ten different kinds of personal data that are related to daily life aspects. Interestingly, there are 3 papers that stated special kind of personal data, visualizing moments by posting memory photos on a timeline \cite{r21}, smartphones usage \cite{r11}, tracking Geo location \cite{r16} and driving habits \cite{r16}. 
By summarising the findings, there are 14 kinds of information that have been listed as personal data: Body information, health, diet, fitness,  physical activity, social connectedness, visual moments, sleep quality, financial aspects, smartphone usage, psychological data and traits, productivity, lifelogging, environmental property statue, energy consumption and driving habits. Among these personal data, health condition was the most data that have been studied. Seven studies have focused on this kind of personal data as tracked data using Desktop or mobile applications. For instance, 2 studies focused on diabetes tracking \cite{r7}, \cite{r8}, 1 study focused on chronic disease students to help them moving to adult-oriented health care\cite{r9}, 2 studies focused on maintaining good sleep pattern\cite{r13}, \cite{r29} and 2 studies involved Diet and fitness along with other kinds of tracked personal data \cite{r5}, \cite{r17}.

\subsection{ Type of Users}
\label{TU}
Based on the papers found, Table 5 represents that 15 papers addressed the second research question. We grouped the users into two groups. Firstly, the public who are interested in self-tracking. Secondly, users who have been recruited to evaluate an application or framework. There are two attributes considered of the target audience in these papers: users' age and experience level.

In terms of the first group, six papers covered interested public participants to explore their interaction, engagement and feedback on self-tracking applications or prototypes. In addition, two papers \cite{r5}, \cite{r6} investigated public users' -who have been encouraged by the quantified self-movement that was established in 2008- interaction with personal data collected and feedback on application reviews. In these studies, users recorded their experience in collecting their data, stating the data items been tracked, listing the tools been used to collect these data, explaining the benefits of tracking personal data, and exploring the UX issues when using self-tracking tools. 
Another paper covered public people who are interested in tracking specific kinds of personal data such as health and physical activity \cite{r22}. Study \cite{r23} explored the effectiveness of elicitation interviews to explore users experience in collecting and visualising data by interviewing ten public individuals who are interested in data visualisation. Study \cite{r24} involved people who were motivated to self-track to find out the effectiveness of building their tracking activity. Finally, the study \cite{r29} focused on developing an application that supports users' sleeping patterns. The participants involved evaluated the application based on its learnability, appeal, and distraction.

In terms of the second group, two papers targeted diabetes people and educators: One paper targeted diabetes people who do not know self-tracking habits to discover the effectiveness of data visualisation adopted in diabetes commercial applications  \cite{r7}. The other study targeted diabetes people who have been trained to track their health data in order to understand how these groups of people are tracking their habits for health improvement \cite{r8}. One paper recruited students to act as designers to specify the suitable visualisation interface \cite{r9}. Another paper recruited individuals from the HCI lab to conduct the study of evaluating smartphone usage data visualisation \cite{r11}. Another couple of papers recruited a naive individual to understand users needs and expectations concerning personalising data visualisation \cite{r18}, \cite{r26}. There is one paper that involved a group of developers who are interested in building self-tracking tools in order to track the mood\cite{r16}. Finally, two papers have not mentioned the users' types involved in the study. These studies focused on finding out the guidelines and recommendations in designing visualisation interfaces on tablets, coping challenges of visualising big data on mobile devices  \cite{r3},\cite{r4}.   

According to the focus points that we have decided, only six have indicated the users' age ranges from 22 to above 50 years. Furthermore, only four papers have indicated the knowledge level of the users that ranges from naïve to expert users. 
Although six papers have focused on public users, these studies focused on types of tracked personal data but not visualising these data. Therefore, more studies are needed on how public users interact with visualising their data.

\subsection{Mobile Devices: Benefits and Challenges }
\label{Mconspros}
To address the third research question, we found 15 papers covering various types of mobile devices adopted in self-tracking personal data, the advantages of using mobile devices and the challenges of mobile devices in terms of hardware and data visualisation. 
As shown in Table 6, out of 15 papers, 13 studies have specified the types of devices included in the study. 
We classified these papers into two groups, the former is related to papers that targeted smartphone applications in terms of building and evaluating them, and the latter is related to combining smartphone and wearable devices.

To elaborate on the first group, We found eight studies focused on developing tracking applications that suit the currently available mobile devices' operating systems. One paper involved a group of developers who are interested in building self-tracking tools in order to track the mood and explored the interface issues when adopting data visualisation on tablets \cite{r3}. Another study included creating an application on mobile devices with different Operating systems\cite{r4}. Furthermore, we found study evaluated diabetes commercial application and its data visualisation effectiveness to deliver information \cite{r7} and a study evaluated small multiples and animation data visualisation on a couple of operating system (Android and ISO) \cite{r10}. We found a study focused on smartphone usage by building a prototype of a person analytics tool to get insights on personal life through smartphone usage\cite{r11}. Similar to the previously mentioned study, there is another paper discussed how smartphone usage affects sleeping habits by approaching three services: triggering for more walk, smartphone time usage and sleeping duration and relaxing before bedtime \cite{r13}. In addition to these studies, there is another study that helped people to sleep better by developing applications \cite{r29}. The latest study focused on evaluating the usability of 62 applications in order to build an application that enables users to customise their tracking activity and study\cite{r24}.

In terms of the second group, we found that Samsung smartwatch and Fitbit were involved in the following study as personal data collection sources. For instance, study \cite{r18} adopted wearable devices as a personal data source. The application built in this study has used these wearable devices to collect personal data and visualise it using timeline and colours to enable users to see their photos at different moments. Study \cite{r25} adopted a Fitbit bracelet to collect and visualise data doubles. Study \cite{r26} familiarised mobile applications and bracelets to enhance and explore users experiences to collect and visualise their data, and study \cite{r20} introduced smartwatch as a controller when visualising data on a large screen. Study \cite{r19} presented various types of wearable devices such as Samsung smartwatch, Fitbit HR, spire devise and mood metric ring.
There are 3 papers that have not mention the type of devices ( \cite{r6} , \cite{r8} , \cite{r9})

There are indeed various studies that cover smartphones and tablets in different areas. These areas included developing personal data applications for different purposes, evaluating and reviewing current self-tracking applications, and exploring UX issues in data visualisation. Nevertheless, there is limited focus on the device performance and capability to adopt personal data visualisation.

To summarise the findings of this section, smartphones, tablets, and smartwatches or bracelets are identified as mobile devices. While mobile devices as device have been mentioned clearly in 4 studies \cite{r4}\cite{r10}\cite{r29}\cite{r24}, smartphone as a general term has been clearly identified in three papers \cite{r7}\cite{r11} \cite {r19} . Tablets have been identified clearly in one study \cite{r3}. Interestingly, smartwatches or bracelets have been included in five papers as a wearable devices that can be used to track physical activity as personal data. 

Table 7 has shown that six studies mentioned the benefits of adopting mobile devices, and another four mentioned mobile devices' issues when visualising personal data and using self-tracking applications.

We found that, small screen is the main issues of adopting personal data visualisation on mobile devices \cite{r7}\cite{r22}\cite{r24}. Furthermore, scale-ability \cite{r10}, battery drain \cite{r24} and limited functionality issues were results of studies conducted in reviewing applications \cite{r7} .

In contrast, the built in sensors\cite{r16}, powerful processor\cite{r10}, high resolution display \cite{r10}, providing personal space \cite{r9}, low maintenance\cite{r22} and easy access to data \cite{r2}\cite{r22} have supported mobile devices to be adopted in visualizing personal data. 

\subsection{Visualisation Tasks and Techniques}
\label{vstaskatech}
To address the fourth research question, we classify visualisation features into three general aspects: visualisation tasks, visualisation types, and interaction techniques. 
The first aspect is visualisation tasks which are shown in table 9. We found 14 papers that have covered visualisation tasks. These tasks are either adopted or need to be adopted to get better visualisation interaction. The second aspect is visualisation types which are shown in table 8. Nine studies have evaluated various visualisation types either directly or as part of a self-tracking application. Finally, the third aspect is related to visualisation interaction techniques, in which four papers have suggested that there is a need for further visualisation interaction techniques.

The adopted visualisation tasks are read-only, in which users can interact with their self-tracking application by reading the available graphs. For example, a spatial map is utilised to show users photos based on a timeline and enable them to remember facts that happened on a specific day \cite {r6}. There is a limited number of tasks attempted to interact with personal information such as sharing information \cite{r16},\cite{r21}, adjusting visualisation \cite{r21}, filtering \cite{r28}, \cite{r20} , elaborating \cite{r21} , selecting \cite{r21}, and configuring data \cite{r21}. Furthermore, identify and compare trends \cite{r10} adopted to compare between two visualisation types performance in mobile devices. On the other hand, some tasks have been required to be adopted, which are building visualisation \cite{r29} and setting self-tracking goals by oneself \cite{r27}. 

Regarding the visualisation types that have been mentioned in the papers. All of them were 2D charts, line graph \cite{r5}\cite{r6}, pie charts \cite{r7}, info-graphic visualisation \cite{r14} \cite{r12}, map visualisation \cite{r21}, colours\cite{r23} \cite{r9}, patterns \cite{r23} , time-series graph \cite{r6}, Multivariate and animation visualisation \cite{r10}. 

In terms of the visualisation techniques, we found four different techniques that need to be explored. The first is data storytelling, which emerged a new class of self-service application, and include dynamic query, visual analytics and interactive presentation\cite{r12}. Furthermore, the same source opened a new research area related to representing data physically, in which data can be represented in 5 forms of physical modalities such as data sculptures, Ambient display, pixel sculpture, object augmentation, wearable visualisation  \cite{r34}.  

The other two techniques that are needed to support understanding visualisation are adding natural language to give a description of the presented visualisation types and avoid various understanding \cite{r27}, \cite{r26}, and auto-generated visualisation \cite{r24}. From our point of view, there are some missing techniques to be adopted to visualise personal data on mobile devices such as augmented reality. As the current smartphones have already adopt augmented reality, it would be valuable and exciting to adopt personal data visualisation using this technique. 

\subsection{Feedback}
\label{fb}

We found only eight papers that have included users reviews of data visualisation that have been adopted in self-tracking applications. There is one study that has focused on building an application but, it has not been reviewed yet\cite{r9}.

These eight papers have summarised users feedback in which we can discuss them from two perspectives. Firstly, there are three papers involved in reviewing used visualisation types and users engagement. In study \cite{r6}, the researcher collected information from the quantified-self website in which public users posted their experiences and shared knowledge related to UX issues when using Personal informatics applications. To summarise the results of this study, the researcher has collected the users' feedback and found that line graph is not easy to be understood by everyone, time series as a graph is the most used visualisation type as it is easy to track various progressing. In the quantified self-study that collected information from user experience and shared knowledge, it is noticed that not all users can understand the simple graphs such as line graphs and most of the users liked the sharing feature \cite{r5}. In the same study, users developed some suggestions, including fun factors such as rewards and budges when they achieve goals to increase the level of engagement and simplify user inputs.
Another study showed different user feedback compared two visualisation types  (small multiples and animation visualisation). Users preferred small multiples scatterplots because it was faster to find information than the animation. However, the group that experienced animation visualisation reported more confidence than the first group in terms of confidence. It can be noticed that both visualisations suited mobile devices, with large datasets cite{r10}. Further studies need to be done to check the compatibility of various visualisation types on mobile devices. 

Secondly, four papers summarised the recommendation that users stated in the application/prototype that they reviewed. 2 of them have supported the need for natural language to be integrated with data visualisation to have better understanding \cite{r26}, \cite{r27}. The context of both studies was distinguished; in the study, \cite{r26} 14 naive participants were recruited to share their experience in collecting their data and visualised it. Therefore, they have suggested that adding natural language with data visualisation would improve their engagement and understanding. In study, \cite{r27} researcher designed a hypothesis for better data visualisation understanding. Thus, they suggested adding natural language to explain the presented data visualisation so users get a similar understanding of the presented data. Based on the feedback collected from quantified selves in study \cite{r6}, fun factors need to be considered in designing personal data visualisation applications as it encourages users to be engaged with their data. These fun factors can be getting budges for goal achievement or enabling sharing their achievements with others. In paper \cite{r7}, 13 diabetes users reviewed three commercial applications and filled a survey that is related to their interaction with these applications. They stated that data visualisation was not offering the help of understanding the information, not only showing numbers, but they want something that graphically tells them what they do is right or wrong, they want to know how their choice would affect their sugar level. In the same study, the pie chart took the high rates in terms of being easy to read as users claimed that it helped in gaining cause and effect concept. There were some recommendations as well. That there is a need for natural language to support data visualisation interpretation. 
In the activity tracking technology study, the participants preferred websites over mobile devices as they showed graphs with more details. Although this research paper had multiple screenshots that presented data visualisation graphs, the research did not focus on this aspect when reviewing the application  \cite{r22}. Although testing visualisation was not an aspect of being examined in the field study for OMNI application, participants commented actively on the visualisation aspect. Therefore,  researchers agreed that the application needs better data visualisation  \cite{r24}. 
In the smartphone usage study, the interviewed volunteers were asked about three issues, usability, visual design and other suggestions. In terms of usability, most of the volunteers recorded positive responses to learning more about themselves using the tool. In addition, they were interested in exploring their smartphone usage data. In terms of visual design,  most users presented a positive attitude towards the interface in general. The flexibility, friendly and smooth interaction with the application make users confident to use the application. In contrast, there were a couple of issues related to the same aspect. For instance, colour coding was confusing for some users, and other users claim that the application was limited in terms of further exploring data \cite{r11}.

\section{Discussion}
\subsection{Self-tracked personal data}
Based on the above section, we think that the analysed studies have perfectly answered the research question that we proposed related to the types of tracked personal data. However, the only personal data examined and attempted in different studies were limited. For example, diabetes, smartphone usage, mood and sleeping pattern were the personal data tracked using mobile applications. 
We expect this review to be used as a start for researchers to acknowledge the various types of personal data individuals would like to track. Furthermore, it opens another research area in investigating more personal data to be tracked and visualised, especially with the emergence of mobile devices and wearable devices and their built-in sensors.

\subsection{Type of Users }
In terms of types of users, we observed that participants involved in the studies were varied. Most of them have been chosen based on specific inclusion criteria that suited the study target and type or recruited from the same team that was developing the project. Researchers may have guidance on various types of users and participants that would be recruited. However, from our point of view, we would prefer to focus more on non-expert individuals of various ages and levels of experiences to explore their interaction and interest to interact with personal data visualisation. Furthermore, we would focus on mobile device users to understand how frequently they are using self-tracking applications and how they interact with data visualisation presented in such applications. 

\subsection{Mobile Devices, Benefits and Challenges}
We found that smartphones as mobile devices were adopted in various studies. However, limited papers have focused on the benefits and challenges of adopting data visualisation in mobile devices. Furthermore, the available papers indeed specified the type of mobile device used in the study, yet, few of them specified the capability of these devices to adopt data visualisation. Thus, we need more research on mobile phones' capability to adopt data visualisation and specify visualisation types, tasks, and techniques that could be adopted using mobile devices. 
\subsection{Visualisation Tasks and Technique}
The role of data visualisation to present tracking personal data is important. Therefore, we discussed visualisation tasks, techniques, and types mentioned amongst the reviewed papers in this section. 
\subsubsection{Visualisation Tasks:}
Although there are some visualisation tasks mentioned in various papers, we recommend that a large study is needed to examine these tasks and its adoption to mobile devices environment.Furthermore, we recommend  that it is needed to compare these tasks with the ten frequent tasks that Saket et al. published in their paper\cite{r40} to have a full list of basic visualisation types and analytic tasks that are required to analyse personal data. 
\subsubsection{Visualisation Types:}

We recommend to do more research on suitable visualisation charts that match users preferences as visualisation types in the result section do not sufficiently address the data visualisation research question. As there are still missing visualisation types that have not been mentioned across the selected studies. We expect that it would be useful to investigate how to select the best visualisation type to present the needed data and the visualisation type that might suit various types of personal data.
\subsubsection{Visualisation Interaction Techniques:}

We found a limited number of papers that mentioned visualisation interaction techniques, and all of them are suggestions from users feedback. The suggested techniques were adding natural language to interpret the information represented in the visualisation type and auto-generation visualisation types to better select the best visualisation to present personal data. 
We think that this part requires more research studies, as some techniques have not been investigated yet. For example, augmented reality, in which charts can be represented as 3D charts. Augmented reality has been applied already to present big data in different sectors. We are suggesting to examine its role in visualising personal data \cite {r34}.

\subsection{Feedback}
 Three interesting recommendations came out of the studies found, such as adopting natural language to support understanding the presented visualisation, adding budges for completing goals, and auto-generating for visualisation types as it would better select the suitable visualisation type. Yet, from our point of view, the available feedback from users regarding data visualisations is not sufficient. The focus of the papers we found was to evaluate self-tracking applications in general. However, one paper mentioned that participants did evaluate data visualisation, although that was not part of the application evaluation. Therefore, we suggest exploring personal data visualisation on mobile devices and how people are interacting with it is needed to understand the public's requirement in terms of visualising personal data.

The limitations of our SLR are, most of the papers found focused on self-tracking aspects. Data visualisation as part of self-tracking were discussed as a minor aspect of visualisation types, tasks and interface design. Furthermore, although many papers found mentioned mobile applications, none of them focused mainly on the benefits and challenges of adopting personal data visualisation on mobile devices. Therefore, this information is limited, and research needs to be conducted to summarise the best practice of personal data visualisation using mobile devices. A further limitation is the found user reviews of self-tracking applications, we tried to find out the feedback related to data visualisation, but it was limited as the focus was reviewing the entire application.

\section{Conclusion}
\label{con}
To summarise this SLR, the papers that have been selected are based on the inclusion criteria provided and the search phrase adopted. Then, the researcher summarised the found papers and found out the results for the research questions. Many papers have addressed the first research questions related to the type of tracked personal data. However, the found papers have not sufficiently addressed the other research questions. Therefore, to bridge this gap, more studies need to be done in the following areas: exploring the public's attitude toward visualising personal data and engaging with it. Another research area that needs to be focused on is the capability of mobile devices to adopt various visualisation techniques and identify the basic and advanced tasks that users can do using data visualisation.


%
%

\newpage
\appendix

\begin{table}
\caption{A list of papers}
\label{tab:12A}       
\begin{tabular}{ll}
\hline\noalign{\smallskip}
 Title&coding\\
\noalign{\smallskip}\hline\noalign{\smallskip}
Reaching Broader Audiences With Data Visualization & P1\\
An evaluation-guided approach for effective data visualization on tablets & P2\\
Making Sense: An Innovative Data Visualization Application Utilized Via Mobile Platform & P3\\
Understanding quantified-selfers' practices in collecting and exploring personal data & P4 \\
Exploring UX issues in Quantified Self technologies & P5 \\
Questioning the Reflection Paradigm for Diabetes Mobile Apps & P6\\
Personal discovery in diabetes self-management: discovering cause and effect using self-monitoring data & P7\\
The KOOLO app & P8\\
A comparative evaluation of animation and small multiples for trend visualization on mobile phones & P9\\
A personal visual analytic on smartphone usage data & P10 \\
Civic Participation and Empowerment through Visualization & P11\\
Nudge Better Quantified-Self with Context-Aware and Proactive Services & P12 \\
The quantified self: Fundamental disruption in big data science and biological discovery & P13\\
A quantum of self: A study of self-quantification and self-disclosure & P14 \\
The diverse domains of quantified selves: self-tracking modes and data veillance & P15\\
Regaining control–the “quantified self” movement and the creation of the postmodern human & P16 \\
Envisioning the future of personalization through personal informatics: A user study & P17 \\
Quantified factory worker-expert evaluation and ethical considerations of wearable self-tracking devices & P18\\
When david meets goliath: Combining smartwatches with a large vertical display for visual data exploration & P19\\
Visual mementos: Reflecting memories with personal data & P20\\
Revisiting personal information management through information practices with activity tracking technology & P21\\
The elicitation interview technique: Capturing people's experiences of data representations & P22\\
OmniTrack: A flexible self-tracking approach leveraging semi-automated tracking & P23\\
Visualized and interacted life: Personal analytics and engagements with data doubles & P24\\
Designing a personal informatics system for users without experience in self-tracking: a case study &P25\\
Know thyself: a theory of the self for personal informatics & P26\\
Personal visualization and personal visual analytics & P27\\
ShutEye: encouraging awareness of healthy sleep recommendations with a mobile, peripheral display & P28\\
Expanding research methods for a realistic understanding of personal visualization & P29\\

\noalign{\smallskip}\hline
\end{tabular}
\end{table}

%
%


\

\bibliographystyle{unsrt}  
\bibliography{ReferenceList}

\end{document}